\documentclass[journal]{IEEEtran}
\usepackage{amsmath,amssymb,amsfonts,mathtools}
\usepackage{algorithm}
\usepackage{algpseudocode}
\usepackage{graphicx}
\usepackage{array}
\usepackage{stfloats}
\usepackage{cite}
\usepackage{url}
\usepackage{textcomp}
\usepackage{booktabs}
\usepackage{tabularx}
\usepackage{multirow}
\usepackage{xcolor}

\definecolor{wheat}{rgb}{0.96,0.87,0.70}

\definecolor{mario}{rgb}{0.8,0.8,1}

\definecolor{filippo}{rgb}{1,0.8,0.1}

\begin{document}
\bstctlcite{IEEEexample:BSTcontrol}
\title{A Heuristic for Scheduling Autonomous Vehicles at Intersections}
\author{Saugat Shahi, Filippo Fabiani and Mario Zanon}

\maketitle

\begin{abstract}
We consider the coordination of autonomous vehicles at unsignalized intersections, where trajectory planning and crossing-order decisions are intrinsically coupled, thereby making their simultaneous optimization intractable in practice. To this end, we propose a heuristic based on a hierarchical structure consisting of a series of nonlinear programs (NLP) in the lower level, which optimizes trajectories of each vehicle for a given crossing order, and a high-level heuristic to design new crossing schedules that exploits sensitivity information from the lower-level NLPs. Extensive numerical simulations across a broad range of traffic scenarios demonstrate that our strategy achieves substantially improved scheduling with respect to first-in-first-out ordering while preserving computational tractability.
\end{abstract}

\section{Introduction}
The coordination of autonomous vehicles at unsignalized intersections is a crucial problem in transportation systems of the future, where the replacement of traffic rules with optimized coordination promises significant improvements in terms of travel time, fuel consumption, and throughput~\cite{fabiani2019multi, levin2017conflict, hult2019optimal}. This problem requires one to simultaneously perform (a) trajectory planning, and (b) collision-free ordering of vehicles.

Existing approaches address this challenge by using different strategies, see, e.g., the survey papers~\cite{namazi2019intelligent, wu2022intersection, rios2016survey,al2022signalized}.  In particular, mixed-integer formulations~\cite{fayazi2018mixed,muller2016intersection,fayazi2017optimal,yu2018integrated} encode the crossing order through binary variables and optimize it jointly with the trajectories. While such formulations capture the coupling between order and motion explicitly, the number of binary variables however grows combinatorially with the number of vehicles, thereby hindering scalability to realistic scenarios. 

To alleviate the complexity of mixed-integer approaches, some techniques resort to heuristics in order to decide the crossing order.
For instance, rule-based policies fix the order through predefined priorities, e.g., derived from right-of-way conventions~\cite{qian2014priority} or from the kinematic state of each approaching vehicle~\cite{zhang2015state}. Reservation-based schemes, instead, grant conflict-free space-time slots on a first-come basis through a central manager~\cite{dresner2008multiagent, levin2016optimizing}. These approaches are computationally light and provably collision-free, but they decide the crossing order from simple fixed criteria that do not account for vehicle dynamics or the individual cost of each vehicle. Consequently, these heuristics may perform very poorly in scenarios where the vehicles have heterogeneous costs.

The formulation we adopt builds upon the work of~\cite{Hult2016PrimalDecomp,Zanon2017Asynchronous,hult2019optimal, Hult2018ECC,Hult2020Coordination,Hult2020Experimental,Hult2022Intersection}, which casts intersection coordination as a constrained optimal control problem and excludes collisions through precedence constraints based on the times at which vehicles enter and leave the intersection. This approach is flexible, as it offers the opportunity to include energy-optimal objectives~\cite{zhang2016optimal, Hult2018ECC}, vehicles that turn inside the intersection~\cite{Hult2019ECC}, as well as networks of adjacent intersections~\cite{chalaki2021optimal}. Closely related formulations are the convex conflict-resolution model of~\cite{Murgovski2015Convex}, the cooperative receding-horizon scheme of~\cite{deCampos2014Cooperative}, the safety-and-liveness framework of~\cite{KimKumar2014}, and the parallelized distributed MPC of~\cite{Katriniok2017Distributed}. Nevertheless, these formulations all share the same limitation, namely they assume a fixed crossing order, decoupling the scheduling decision from motion planning and leaving part of the achievable performance unexplored.

Decoupling scheduling and motion planning motivates the interest in schedule-refinement heuristics, which attempt to improve an initial schedule based on some information given by the motion planning task. In~\cite{Hult2018CDC}, the order is selected by solving a mixed-integer quadratic approximation of the coordination problem, and in~\cite{Scheffe2022DynamicPriority} vehicle priorities are reassigned within prioritized distributed model predictive control.

Unlike available methods, we propose a coordination scheme that refines the crossing order using sensitivity information from the motion planner. For a fixed order, a single NLP jointly optimizes the trajectories and the entry and exit times of all vehicles. A higher-level heuristic then refines this order using the dual variables of the NLP: the Lagrange multipliers of the precedence constraints are the first-order sensitivities of the cost associated with each ordering decision.
Since this information is local, each candidate swap is retained only if the cost is actually reduced. The resulting heuristic explores only the most promising part of the permutation space and solves a handful of NLPs, in contrast to  schemes based on integer variables, whose computational cost grows combinatorially with the number of vehicles. Extensive experiments over a broad set of traffic scenarios show that our approach recovers most of the optimality gap of first-in-first-out (FIFO) scheduling. The extension of this approach to make it suitable for dynamic environments in which vehicles keep arriving at the intersection and leaving it is the subject of ongoing research.
%, making it suitable to real-time intersection coordination.

\smallskip

\subsubsection*{Notation}
The set of consecutive integers from $a$ to $b$ is denoted $\mathbb{I}_a^b \coloneqq \{a, a+1, \ldots, b\}$, with $\mathbb{I}_a^b = \emptyset$ if $a > b$. For a finite set $\mathcal{S}$, $|\mathcal{S}|$ denotes its cardinality. Throughout the paper, $M$ is the number of vehicles and $N$ the number of time  steps, with sampling time $t_\mathrm{s}$, and the superscript $\dagger \in \{\mathrm{in}, \mathrm{out}\}$ is a placeholder for entry and exit quantities, so that any statement involving $\dagger$ is understood to hold for both values. A \emph{schedule} $S = (S_1, \ldots, S_M)$ is a permutation of $\mathbb{I}_1^M$, where $S_c \in \mathbb{I}_1^M$ is the index of the vehicle occupying the position $c$ in the crossing order. For each vehicle $i$, the state and input trajectories are collected as $x_i \coloneqq (x_{i,0}, \ldots, x_{i,N})$ and $u_i \coloneqq (u_{i,0}, \ldots, u_{i,N-1})$, and the corresponding quantities for all vehicles as $x \coloneqq (x_1, \ldots, x_M)$, with $u$ and $t$ defined analogously, $t_i \coloneqq (t_i^\mathrm{in},t_i^\mathrm{out})$. Superscript $(r)$ denotes quantities at iteration $r$ of the schedule-refinement heuristic, while $\cdot^\star$ denotes optimal quantities.
 
\begin{figure}[t]
	\centering
	\includegraphics[width=0.8\linewidth]{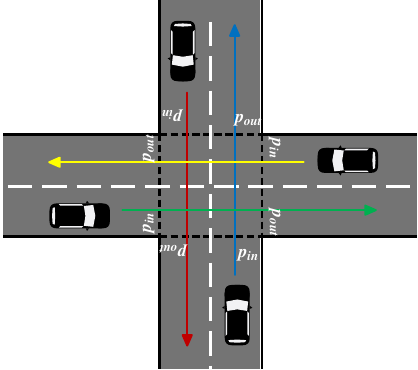}
	\caption{Scenario considered in this paper. Vehicles approach a four-way unsignalized intersection from different directions. Colored arrows show the fixed straight paths, while the dashed black square marks the conflict zone where vehicles interact. Finally, \(p_i^{\mathrm{in}}\) and \(p_i^{\mathrm{out}}\) are the entry and exit positions.}
	\label{fig:intersection_diagram}
\end{figure}

\section{Problem formulation}
We aim at solving the problem of coordinating $M$ cooperative autonomous vehicles approaching an unsignalized intersection, as illustrated in Figure~\ref{fig:intersection_diagram}. The main goal is to compute an optimal crossing order (schedule) with corresponding optimal state and input trajectories for each vehicle $i$. To this end, we implicitly assume that each vehicle follows a predefined path, and for simplicity no vehicle changes lane. 

At time $k\in\mathbb{I}_{0}^{N}$, each vehicle $i$ has state $x_{i,k}\in\mathbb{R}^{n_{x_i}}$ and control input $u_{i,k}\in\mathbb{R}^{n_{u_i}}$, with dynamics evolving as:
% We denote the vehicle dynamics as
\begin{equation}
x_{i,k+1} = f_i(x_{i,k}, u_{i,k}),
\label{eq:dynamics_function}
\end{equation}
which we assume to be twice continuously differentiable.
We formulate the local constraints imposed at each $k$ as
\begin{equation}
h_i(x_{i,k}, u_{i,k}) \le 0.
\label{eq:path_constraints}
\end{equation}
These constraints include, for instance, actuator limitation and lane keeping. 
Each vehicle \(i\) enters and exits the intersection at specific positions along its path, denoted by \(p_i^{\mathrm{in}}\) and \(p_i^{\mathrm{out}}\), respectively. We define the corresponding entry and exit times \(t_i^{\mathrm{in}}\) and \(t_i^{\mathrm{out}}\) implicitly by exploiting functions $p_i(t,x_i,u_i)$, defining the position at any time $t$, as follows:
\begin{equation}
p_i(t_i^{\dagger},x_i,u_i) = p_i^{\dagger}, \qquad \dagger \in \{\mathrm{in},\mathrm{out}\}.
\label{eq:entry_exit_times}
\end{equation}

To ensure collision-free operation of the vehicles in the intersection, we impose precedence constraints~\cite{hult2019optimal, Hult2018CDC}. These specifically enforce that at most one vehicle occupies the intersection at any time $t$, i.e., if vehicle \(i\) is scheduled to pass through the intersection before vehicle \(j\), then 
\begin{equation}
t_j^{\mathrm{in}} \geq t_i^{\mathrm{out}}.
\label{eq:precedence_pairwise}
\end{equation}

The optimization variables are therefore the states $x\coloneqq (x_1,\ldots,x_M)$, controls $u \coloneqq (u_1,\ldots,u_M)$, entry/exit times $t \coloneqq (t_1,\ldots,t_M)$ of all vehicles,

where $t_i \coloneqq (t_i^{\mathrm{in}},t_i^{\mathrm{out}})$, as well as the schedule $S:=(S_1,\ldots,S_M)$.
The overall cost function is given by the sum of individual cost functions associated with each vehicle, i.e.,
\begin{align*}
    J(x,u) &\coloneqq \sum_{i=1}^M J_i(x_i,u_i), \\ %\text{ with }\\
    J_i(x_i,u_i) &\coloneqq \sum_{k=0}^{N-1}
		\ell_i(x_{i,k}, u_{i,k}) + \ell_{i,\mathrm{f}}(x_{i,N}),
\end{align*}
where $\ell_i$ and $\ell_{i,\mathrm{f}}$ are, respectively, the local stage and terminal costs.
The overall coordination problem can therefore be formulated as the following optimization problem:
\begin{align}
    \label{eq:coordination_ocp_mi}
    \min_S \ V(S),
\end{align}
where, given initial states $\hat{x}_{i,0}$, we define %$V$ as
\begin{subequations}
\label{eq:coordination_ocp}
\begin{alignat}{2}
    V(S)\coloneqq\min_{x,u,t} \ 
    & J(x,u)
    \label{eq:objective_function} \\
    \mathrm{s.t.} \ 
    & x_{i,0} = \hat{x}_{i,0},
    && \forall \, i \in \mathbb{I}_1^M,
    \label{eq:init_conditions} \\
    & x_{i,k+1} = f_i(x_{i,k}, u_{i,k}),
    && \forall\, i \in \mathbb{I}_1^M,\, k \in \mathbb{I}_0^{N-1},
    \label{eq:dynamics_compact} \\
    & h_i(x_{i,k}, u_{i,k}) \le 0,
    && \forall\, i \in \mathbb{I}_1^M,\, k \in \mathbb{I}_0^{N-1},
    \label{eq:path_constraints_2} \\
    & t_{S_c}^{\mathrm{in}} \ge t_{S_{c-1}}^{\mathrm{out}},
    && \forall\, c \in \mathbb{I}_2^M,
    \label{eq:precedenceconstraint} \\
    & p_i(t_i^{\mathrm{in}},x_i,u_i) = p_i^{\mathrm{in}},
    && \forall\, i \in \mathbb{I}_1^M,
    \label{eq:entry_position} \\
    & p_i(t_i^{\mathrm{out}},x_i,u_i) = p_i^{\mathrm{out}},
    && \forall\, i \in \mathbb{I}_1^M.
    \label{eq:exit_position}
\end{alignat}
\end{subequations}
Problem~\eqref{eq:coordination_ocp} is an NLP that can be solved efficiently, though it does not optimize the schedule $S$, therefore missing one of the original aims, hence the need for Problem~\eqref{eq:coordination_ocp_mi}.
By optimizing over $S$, however, \eqref{eq:coordination_ocp_mi} is a mixed-integer nonlinear program (MINLP), which is typically prohibitive to solve in real time, unless $M$ is very small. A viable approach to circumvent this issue is to separate the schedule computation from the rest of the coordination problem, and accept that some degree of suboptimality might thus be introduced. This approach has been followed in several works on optimization-based coordination of autonomous vehicles at intersections, including centralized, semi-distributed, experimentally validated, and heuristic formulations, see, e.g.,~\cite{hult2019optimal,Hult2022Intersection,Hult2020Coordination,Hult2020Experimental,Hult2019ECC,Hult2018CDC,Hult2018ECC,Faris2024TCST}. 
We follow a similar approach by formulating the problem in a hierarchical structure: the lower level solves~\eqref{eq:coordination_ocp} for a fixed schedule $S$, and a higher-level heuristic generates new schedules that reduce the overall cost, as detailed next.

\section{Schedule-refining heuristic}\label{sec:heuristic}
A popular heuristic consists in fixing the schedule based on a FIFO approach, i.e., by ordering the vehicles based on their arrival time in the absence of the other vehicles. This approach is very simple, computationally inexpensive, and would arguably deliver a very good solution in case $J_i=J_j$ for all $i,j$. However, in the general case, the costs differ, and this approach can yield schedules that are far from optimal, as we will also illustrate later in our numerical simulations.

An alternative heuristic solves a quadratic approximation of~\eqref{eq:coordination_ocp_mi} to compute a schedule~\cite{Hult2018CDC}. Even though solvers for quadratic mixed-integer programs are becoming increasingly efficient, the problems they solve are combinatorial and still do not scale well as the number of vehicles becomes large.

We therefore propose a heuristic based on local information about optimality while avoiding computationally demanding algorithms. To this end, we exploit parametric sensitivities to iteratively improve a given schedule as follows. For a given \( S \), let \( \lambda_c \ge 0\) be the Lagrange multiplier associated with the constraint
\[
t_{S_{c}}^{\mathrm{out}} - t_{S_{c+1}}^{\mathrm{in}} \le 0,
\]
i.e., the Lagrangian of~\eqref{eq:coordination_ocp} contains the term \( \lambda_c (t_{S_{c}}^{\mathrm{out}} - t_{S_{c+1}}^{\mathrm{in}}) \).
This multiplier coincides with the first-order sensitivity of the optimal cost with respect to perturbations of the corresponding constraint~\eqref{eq:precedenceconstraint}---see, e.g.,~\cite{nocedal2006numerical,fletcher1987practical}.

In particular, relaxing the constraint to \( t_{S_{c}}^{\mathrm{out}} - t_{S_{c+1}}^{\mathrm{in}} \) decreases the optimal cost at rate \(\lambda_c \). Hence, a larger value of \(\lambda_c \) indicates that relaxing constraint has a significant impact on the optimal cost.

Denoting by $c^\star$ the index of the largest multiplier, we propose to \emph{swap} the corresponding vehicle pair in schedule $S^{(r)}$, where the superscript indicates the generic $r$-th iteration of the proposed scheme, to define the following schedule:
\begin{align}
    \label{eq:swap}
    S^\mathrm{swap} = (S_1^{(r)},\ldots,S_{c^\star-1}^{(r)},S_{c^\star+1}^{(r)},S_{c^\star}^{(r)},S_{c^\star+2}^{(r)},\ldots,S_M^{(r)}),
\end{align}
with the hope that the new order yields a lower cost.
While this is justified by the interpretation of the multiplier as a shadow cost, such information is local and can notoriously not be trusted for mixed-integer optimization. It therefore follows that each attempted swap needs to be evaluated to verify that an actual cost improvement is obtained.

Starting from an initial schedule \( S^{(0)} \), the method then proceeds iteratively. 
At iteration \( r \), the NLP \eqref{eq:coordination_ocp} is solved for the current schedule \( S^{(r)} \), yielding the optimal cost \( V(S^{(r)}) \) and the corresponding multipliers \( \lambda^{(r)} \). 
A set of admissible indices $c$ is initialized as \( \mathcal{P}^{(r)} = \mathbb{I}_1^{M-1} \).
Among all admissible indices, the pair to be swapped is selected as \( (S^{(r)}_{c^\star}, S^{(r)}_{c^\star+1}) \), with
\begin{align}
    \label{eq:swap_index}
    c^\star = \arg\max_{c\in \mathcal{P}^{(r)}} \lambda^{(r)}_c.
\end{align}
A new schedule \( S^{\mathrm{swap}} \) is obtained by exchanging the positions of vehicles \( S^{(r)}_{c^\star} \) and \( S^{(r)}_{c^\star+1} \), as per~\eqref{eq:swap}. The NLP is then solved once more to compute the updated cost \( V(S^{\mathrm{swap}}) \). The swap is accepted if
\[
V(S^{\mathrm{swap}}) < V(S^{(r)}),
\]
in which case the schedule is updated as \( S^{(r+1)} = S^{\mathrm{swap}} \). Otherwise, the swap is rejected, and the current schedule is retained. Index $c^\star$ is removed from set $\mathcal{P}^{(r)}$ and a new swap is attempted by selecting the pair according to~\eqref{eq:swap_index}. In case the set $\mathcal{P}^{(r)}$ becomes empty, we terminate and return $S=S^{(r)}$.

\begin{algorithm}[t]
\caption{Lagrange-multiplier-based vehicle reordering}
\label{alg:lambda_swap}
\small
\begin{algorithmic}[1]
\State \textbf{Input:} $S_{\text{init}}, J_{\text{init}}, \lambda_{\text{init}}, t_{\text{in,init}}, t_{\text{out,init}}, K_{\max},\alpha$
\State \textbf{Output:} $S_{\text{best}}, J_{\text{best}}, \lambda_{\text{best}}, t_{\text{in,best}}, t_{\text{out,best}}$
\State Initialize $S_{\text{best}} \gets S_{\text{init}},\; J_{\text{best}} \gets J_{\text{init}},\; \lambda_{\text{best}} \gets \lambda_{\text{init}}$
\State $\mathcal{A} \gets \emptyset,\; \mathcal{R} \gets \emptyset,\; r \gets 0$
\While{$r < K_{\max}$}
    \State Solve NLP~\eqref{eq:coordination_ocp} for $S_{\text{best}}$, extract $\lambda_{\text{best}}$
    \If{$\alpha=2$} \Comment{Heuristic 2: revisitable}
        \State $\mathcal{P} \gets \left\{\, c \in \mathbb{I}_1^{M-1} : \left(S_{c}^{(r)}, S_{c+1}^{(r)}\right) \notin \mathcal{R} \,\right\}$
    \Else \Comment{Heuristic 1: frozen}
        \State $\mathcal{P} \gets \left\{\, c \in \mathbb{I}_1^{M-1} : \left(S_{c}^{(r)}, S_{c+1}^{(r)}\right) \notin \mathcal{A} \cup \mathcal{R} \,\right\}$
    \EndIf
    \If{$\mathcal{P} = \emptyset$}
        \State \textbf{break}
    \EndIf
\State Select $c \in \mathcal{P}$ s.t. $\lambda_{c}\geq\lambda_{c'}, \ \forall \, c' \in \mathcal{P}$
    \State $S_{\text{swap}} \gets \mathrm{swap}\left(S_{\text{best}}, c\right)$
    \State Solve NLP~\eqref{eq:coordination_ocp} for $S_{\text{swap}}$, obtain $J_{\text{swap}}$
    \If{$J_{\text{swap}} < J_{\text{best}}$}
        \State Update $S_{\text{best}}, J_{\text{best}}, \lambda_{\text{best}}$
        \If{$\alpha=2$} \State $\mathcal{R} \gets \emptyset$ \EndIf
        \State $\mathcal{A} \gets \mathcal{A} \cup \left\{\left(S_{c+1}^{(r)}, S_{c}^{(r)}\right)\right\}$
    \Else
        \State $\mathcal{R} \gets \mathcal{R} \cup \left\{\left(S_{c}^{(r)}, S_{c+1}^{(r)}\right)\right\}$
    \EndIf
    \State $r \gets r + 1$
\EndWhile
\State \textbf{Return:} best solution
\end{algorithmic}
\end{algorithm}

Our approach is reported in Algorithm~\ref{alg:lambda_swap}, where $K_{\max}$ bounds the number of swap attempts. Let \(\mathcal{P}=\mathbb{I}_1^{M-1}\) denote the set of adjacent swap indices, and sets \(\mathcal{A}\) and \(\mathcal{R}\) collect the vehicle pairs of all accepted and rejected swaps at previous iterates, respectively. In this heuristic, which we call frozen swaps or Heuristic 1, ($\alpha=1$ in Algorithm~\ref{alg:lambda_swap}), at every iterate $r$ the swap of each candidate pair $(S_c^{(r)}, S_{c+1}^{(r)})$ is attempted at most once. If the swap succeeds, then $\mathcal{A}\leftarrow \mathcal{A} \cup \{(S_{c+1}^{(r)},S_{c}^{(r)})\}$, otherwise $\mathcal{R}\leftarrow \mathcal{R} \cup \{(S_{c}^{(r)},S_{c+1}^{(r)})\}$. The admissible set at iteration \(r+1\) then becomes:
\begin{equation}
\mathcal{P}^{(r+1)} = \left\{\, c \in \mathbb{I}_1^{M-1} : \left(S_c^{(r)},S_{c+1}^{(r)}\right)\notin \mathcal{A}\cup\mathcal{R} \,\right\}.
\label{eq:frozenheuristic}
\end{equation}
Consequently, the algorithm will never try again to swap the pair $(S_{c}^{(r)},S_{c+1}^{(r)})$ or $(S_{c+1}^{(r)},S_{c}^{(r)})$, such that fewer swap attempts will be performed. 

Heuristic 1 is motivated by the fact that, after swapping a pair of vehicles, Algorithm~\ref{alg:lambda_swap} might attempt to swap the same pair once more, which is not expected to yield any improvement, based on past decisions. Moreover, if at some previous step the algorithm tried a swap in the order of a given pair of vehicles which was rejected, it is likely to try to perform that swap once more at later iterations, as the Lagrange multiplier is expected to still be large. This situation is also unlikely to yield any improvement.
 
For comparison, we also consider a second heuristic, corresponding to $\alpha=2$ in Algorithm~\ref{alg:lambda_swap}, which does not exploit these observations: revisitable swaps, or Heuristic 2, allows any adjacent pair to be revisited at later iterations. A rejected pair is excluded only until a swap is accepted, i.e., the next candidate is the pair with the next-largest multiplier for the same schedule. 
 
Unlike algorithms for mixed-integer optimization, both Heuristic 1 and 2 greedily explore the space of permutations by targeting the pairs with the largest expected impact on the cost. This drastically reduces the computational burden, at the price of obtaining possibly suboptimal solutions. 
These two variants share the same basic mechanism and differ only in the policy used to update the admissible swap set, and their performance will be compared through simulations in Section~\ref{sec:comparative_evaluation}.

\section{Baselines for comparison}\label{sec:MI_formulations}

We propose next two alternative problem formulations that we will use in our simulations as baselines for comparison, so as to evaluate the efficiency and effectiveness of Algorithm~\ref{alg:lambda_swap}.

\subsection{MINLP formulation~\eqref{eq:coordination_ocp_mi}}
An important detail concerning the solution of the MINLP in \eqref{eq:coordination_ocp_mi} is related to the function $p_i$. For a given time $t_i^{\dagger}$, $\dagger\in\{\mathrm{in},\mathrm{out}\}$, one can define $\underline{k}_i^\dagger \coloneqq \max_k$ such that $kt_\mathrm{s}\leq t_i^{\dagger}$,
% \begin{align*}
%     \underline{k}_i^\dagger \coloneqq \max_k \ & k, && \mathrm{s.t.} \ kt_\mathrm{s}\leq t_i^{\dagger},
% \end{align*}
where $t_\mathrm{s}$ is the sampling time. 
Then, function $p_i$ reads as:
\begin{align}
    \label{eq:position_function_k}
p_i(t_i^{\dagger},x_i,u_i)=f_i^p(x_{i,\underline{k}_i^\dagger}^\dagger,u_{i,\underline{k}_i^\dagger}^\dagger,t_i^{\dagger}-\underline{k}_i^\dagger t_\mathrm{s}),
\end{align}
where each $f_i^p$ integrates the vehicle dynamics starting at time $0$ from state $x_{i,\underline{k}_i^\dagger}$ and applying control input $u_{i,\underline{k}_i^\dagger}$ to obtain the position at time $t_i^{\dagger}-\underline{k}_i^\dagger t_\mathrm{s}$.
While some solvers accept constraints formulated in this way (e.g., CasADi~\cite{Andersson2019CasADi} with IPOPT~\cite{wachter2006implementation}, WORHP~\cite{buskens2012esa} or Bonmin~\cite{Bonami2008BONMIN}), some others do not, e.g., Gurobi~\cite{gurobi}. We thus propose an alternative formulation.

To formulate the position constraint more explicitly, we introduce the one-hot encoded binary variable:
\begin{align}
    \label{eq:b_ohe}
    b_{i,k}^\dagger &\in\{0,1\}, && \sum_{k=0}^{N-1} b_{i,k}^\dagger=1,
&& \forall i\in\mathbb{I}_1^M,~\dagger \in\{\mathrm{in},\mathrm{out}\}.
\end{align}
Since we want to exploit this variable to have
\begin{align}
    \label{eq:underline_k_binary}
    \underline{k}_i^\dagger = \sum_{k=0}^{N-1} b_{i,k}^\dagger k, &&  \dagger \in\{\mathrm{in},\mathrm{out}\},
\end{align}
we need to enforce the constraints:
\begin{align}
    \sum_{k=0}^{N-1} b_{i,k}^\dagger k t_\mathrm{s} \le t_i^{\dagger} \le \sum_{k=0}^{N-1} b_{i,k}^\dagger kt_\mathrm{s} + t_\mathrm{s}, &&  \dagger \in\{\mathrm{in},\mathrm{out}\}.
\label{eq:interval_constraint}
\end{align}
We can then exploit~\eqref{eq:position_function_k}, together with the definition of $\underline{k}_i^\dagger$ in~\eqref{eq:underline_k_binary}, to conclude that
\begin{align*}
    x_{i,\underline{k}_i^\dagger}^\dagger &= \sum_{k=0}^{N-1} b_{i,k}^\dagger x_{i,k}, &
    u_{i,\underline{k}_i^\dagger}^\dagger &= \sum_{k=0}^{N-1} b_{i,k}^\dagger u_{i,k}.
\end{align*}

Note that, even for linear vehicle models, \eqref{eq:position_function_k} contains nonlinear terms involving the time $t_i^{\dagger}-\underline{k}_i^\dagger t_\mathrm{s}$. For example, for a double integrator model, \eqref{eq:position_function_k} is quadratic in $t_i^{\dagger}-\underline{k}_i^\dagger t_\mathrm{s}$. 

With $b\coloneqq(b_1,\ldots,b_M)$, $b_i\coloneqq(b_{i,0}^\dagger,\ldots,b_{i,N-1}^\dagger)$, the resulting optimization problem is a MINLP that reads as:

\begin{subequations}
\label{eq:minlp_problem}
\begin{align}
& \min_{x,u,t,S,b} && J(x,u) \\
&~~~\mathrm{ s.t.} && \eqref{eq:init_conditions}\text{--}\eqref{eq:exit_position}, \\
& && \eqref{eq:position_function_k},~\eqref{eq:b_ohe},~\eqref{eq:interval_constraint}.
\label{eq:minlp_position}
\end{align}
\end{subequations}

\subsection{MIQP formulation}
Solving an MINLP is notoriously difficult, thus we propose next a simplified formulation that allows one to turn \eqref{eq:minlp_problem} into a mixed-integer quadratic program (MIQP).

We assume to have a linear vehicle model, such that equality constraints~\eqref{eq:dynamics_compact} and path constraints~\eqref{eq:path_constraints_2} are affine. Then, the only nonconvex constraints are the position constraints~\eqref{eq:entry_position}--\eqref{eq:exit_position}, that implicitly define the entry and exit times of each vehicle. As discussed above, note that also~\eqref{eq:minlp_position} is nonlinear.

To obtain an MIQP formulation, we simplify~\eqref{eq:minlp_position} by discretizing $t_i^{\dagger}$ as described below. For each vehicle $i \in \mathbb{I}_1^M$ and $k \in \mathbb{I}_0^{N-1}$, we introduce the binary variables $b^{\mathrm{in}}_{i,k}, b^{\mathrm{out}}_{i,k} \in \{0,1\}$. To ensure that exactly one time index is selected for each event, we impose the one-hot encoding constraints:
\begin{equation}
\sum_{k=0}^{N-1} b^{\mathrm{in}}_{i,k} = 1, \qquad
\sum_{k=0}^{N-1} b^{\mathrm{out}}_{i,k} = 1,
\qquad \forall i \in \mathbb{I}_1^M.
\label{eq:onehot_bc}
\end{equation}
These variables are used to define the discrete time index $\overline{k}_i^{\dagger} \geq t_i^{\dagger}/t_\mathrm{s}$, selected by $b^{\dagger}_{i,\overline{k}_i^{\dagger}} = 1$, for each $\dagger \in\{\mathrm{in},\mathrm{out}\}$. In other words, $b^{\dagger}$ is the one-hot encoding of the sampling time immediately following the time at which each vehicle enters or exits the intersection. The sampling time immediately preceding the same event, of index $\underline{k}_i^{\dagger} = \overline{k}_i^{\dagger}-1$, does not require a dedicated variable: it is recovered by shifting the index of $b^{\dagger}$, i.e., as $b^{\dagger}_{i,k+1}$. Since $b^{\dagger}$ is one-hot with $\overline{k}_i^{\dagger} \le N-1$, the predecessor index $\overline{k}_i^{\dagger}-1 \in \mathbb{I}_0^{N-2}$ is always well defined. To that end, we impose the following constraints:
\begin{align}
\sum_{k=0}^{N-1} b^{\dagger}_{i,k} \, p_{i,k} &\ge p_i^{\dagger}, && \dagger \in\{\mathrm{in},\mathrm{out}\},
\label{eq:pin_miqp}\\
\sum_{k=0}^{N-1} b^{\dagger}_{i,k+1} \, p_{i,k} &\le p_i^{\dagger}, && \dagger \in\{\mathrm{in},\mathrm{out}\}.
\label{eq:pout_miqp}
\end{align}

The entry and exit times are then linked to the binary variables through
\begin{align}
t_i^{\mathrm{in}} + t_\mathrm{s} &= \sum_{k=0}^{N-1} k t_\mathrm{s} \, b^{\mathrm{in}}_{i,k},
\label{eq:tin_link}\\
t_i^{\mathrm{out}} &= \sum_{k=0}^{N-1} k t_\mathrm{s} \, b^{\mathrm{out}}_{i,k},
\label{eq:tout_link}
\end{align}
for all $i \in \mathbb{I}_1^M$. Equivalently, the entry and exit times are fixed by the selected sample and its predecessor, i.e.,
\begin{equation}
\sum_{k=0}^{N-1} k t_\mathrm{s} \, b^{\dagger}_{i,k+1}
\leq t_i^{\dagger} \leq
\sum_{k=0}^{N-1} k t_\mathrm{s} \, b^{\dagger}_{i,k},
\qquad \dagger \in \{\mathrm{in}, \mathrm{out}\}.
\label{eq:t_interval}
\end{equation}
Note that this formulation is actually a relaxation of the original problem, as 
\begin{align*}
    p_i(t_i^{\mathrm{in}},x_i,u_i) &\geq p_i^{\mathrm{in}}, &
    p_i(t_i^{\mathrm{out}},x_i,u_i) &\leq p_i^{\mathrm{out}},
\end{align*}
such that the obtained solution might actually be infeasible. This issue could be compensated for and is not an issue in the context of this paper, as we only introduce this formulation to have an additional term of comparison that supports the use of our NLP formulation rather than a mixed-integer one.

With the introduced binary variables and linear dynamics, the coordination problem can then be formulated as the MIQP
\begin{subequations}
\label{eq:miqp_problem}
\begin{align}
& \min_{x,u,t,S,b} && J(x,u) \\
& ~~~\mathrm{ s.t.} && \eqref{eq:init_conditions}\text{--}\eqref{eq:precedenceconstraint}, \\
& && \eqref{eq:onehot_bc}\text{--}\eqref{eq:t_interval}.
\end{align}
\end{subequations}

\section{Simulation results}\label{sec:simulation_experiments}
We now evaluate the proposed coordination framework across a range of traffic scenarios, aiming at quantifying both the performance of the computed solution and computational efficiency. We compare our heuristic from Section~\ref{sec:heuristic}, based on NLP~\eqref{eq:coordination_ocp}, with MINLP~\eqref{eq:minlp_problem} and MIQP~\eqref{eq:miqp_problem} of Section~\ref{sec:MI_formulations}.

\subsection{Simulation setup}

All numerical experiments are conducted in a Python-based environment
integrating CasADi, Gurobi, and the NLP solvers IPOPT and WORHP. The NLP
formulations are implemented in CasADi and solved with both IPOPT and
WORHP, and the Lagrange-multiplier-based reordering strategy is likewise
implemented within CasADi. The MIQP and MINLP formulations are solved
using Gurobi.

The parameters used for the simulations are reported in Table~\ref{tab:system_parameters}.
Initial positions and velocities are varied across scenarios to capture heterogeneous traffic conditions, while entry/exit positions are all fixed to \(p_i^{\mathrm{in}} = 0\) m and \(p_i^{\mathrm{out}} = 10\)~m. In the results, vehicles are labeled $0,\ldots,M-1$, and crossing orders are reported as lists of these labels.

\begin{table}[!tbh]
\centering
\caption{Problem parameters}
\label{tab:system_parameters}
\begin{tabularx}{\columnwidth}{@{\extracolsep{\fill}}>{\hsize=0.6\hsize}X >{\hsize=1.6\hsize}X >{\hsize=0.8\hsize}X}
\toprule
\textbf{Parameter} & \textbf{Description} & \textbf{Value} \\
\midrule
$N$               & Number of discrete time steps              & 200 \\
$t_\mathrm{s}$        & Sampling time                              & 0.1 s \\
$M$               & Number of vehicles                         & 3, 6, 9, 12 \\
$p_i^{\mathrm{in}}$ & Entry position of vehicle $i$              & 0 m \\
$p_i^{\mathrm{out}}$& Exit position of vehicle $i$               & 10 m \\
$v_{\mathrm{ref}}$  & Reference velocity                         & 13.88 m/s \\
\bottomrule
\end{tabularx}
\end{table}

The cost function is quadratic and penalizes both the control effort and the deviation of the velocity from the reference value. Specifically, the stage cost for vehicle \(i\in\mathbb{I}_1^M\) reads as:
\begin{align}
    \ell_i(x_{i,k},u_{i,k}) = r_i\,u_{i,k}^2 + q_i\,(v_{i,k}-v_{\mathrm{ref}})^2,
\end{align}
while the terminal cost is defined as
\begin{align}
    \ell_{i,\mathrm{f}}(x_{i,N}) = q_i\,(v_{i,N}-v_{\mathrm{ref}})^2.
\end{align}
The weight \(r_i\) penalizes acceleration and braking effort, while \(q_i\) the velocity-tracking error. Larger values of \(r_i\) discourage aggressive control actions, while larger values of \(q_i\) enforce tighter tracking of the reference speed. These coefficients
hence enable heterogeneous vehicle behavior 
and are selected according to three vehicle classes, as summarized in Table~\ref{tab:vehicle_classes}. 
\begin{table}[!tbh]
\centering
\caption{Vehicle classes and corresponding weights}
\label{tab:vehicle_classes}
\begin{tabular*}{\columnwidth}{@{\extracolsep{\fill}}lcc}
\toprule
\textbf{Vehicle class} & \boldmath{$r_i$} & \boldmath{$q_i$} \\
\midrule
Light  & 1   & 1   \\
Medium & 10  & 10  \\
Heavy  & 100 & 500 \\
\bottomrule
\end{tabular*}
\end{table}

The algorithms are evaluated on unsignalized intersection scenarios with \(M \in \{3,6,9,12\}\) vehicles. For each scenario, four scheduling strategies are then considered:
\begin{itemize}
    \item FIFO scheduling: vehicles ordered according to the entry time they would have in the absence of other vehicles;
    \item Swapping-based heuristic: order refinement, Algorithm~\ref{alg:lambda_swap};
    \item MINLP/MIQP: solution obtained with Gurobi;
    \item Brute-force search: exhaustive enumeration (when computationally feasible).
\end{itemize}

Before proceeding with the schedule optimization, we fix $S$ and compare the different algorithms in
Table~\ref{tab:fifo_solver_comparison}. 
\begin{table*}[!tbh]
\centering
\caption{Comparison of Solvers for FIFO Scheduling}
\label{tab:fifo_solver_comparison}
\begin{tabularx}{\textwidth}{l 
                                >{\centering\arraybackslash}X 
                                >{\centering\arraybackslash}X 
                                >{\centering\arraybackslash}X 
                                >{\centering\arraybackslash}X 
                                >{\centering\arraybackslash}X 
                                >{\centering\arraybackslash}X 
                                >{\centering\arraybackslash}X 
                                >{\centering\arraybackslash}X}
\toprule
Scenario & 
\multicolumn{2}{c}{IPOPT (NLP)} & 
\multicolumn{2}{c}{WORHP (NLP)} & 
\multicolumn{2}{c}{Gurobi (MIQP)} & 
\multicolumn{2}{c}{Gurobi (MINLP)} \\
\cmidrule(lr){2-3} \cmidrule(lr){4-5} \cmidrule(lr){6-7} \cmidrule(lr){8-9}
& Cost & Time (s) 
& Cost & Time (s) 
& Cost & Time (s) 
& Cost & Time (s) \\
\midrule
3 Vehicle  & 2433.84    & 0.25   & 2433.84    & 0.47   & 2546.41    & 387   & 2433.84    & 651.77     \\
6 Vehicle  & 91416.24   & 1.08   & 91416.24   & 3.47   & 113966.95   & 696     & 91416.24   & 3600.00    \\
9 Vehicle  & 1080763.61  & 4.50   & 1080763.61  & 5.62   & 1418502.03  & 1000 & 1080765.10  & 4116.72   \\
12 Vehicle & 4229269.36 & 18.63  & 4229269.36 & 13.90  & 5245213.74 & 1384    & 4229282.94 & 32000.00   \\
\bottomrule
\end{tabularx}
\end{table*}
All solvers except MIQP (which is suboptimal by design) produce the same cost, with deviations below $0.001\%$ of the MINLP solution in the scenarios with $9$ and $12$ vehicles. The NLP solvers, IPOPT and WORHP, are the fastest. IPOPT is generally faster than WORHP on the smaller instances, although the gap narrows as the number of vehicles grows and reverses in the $12$-vehicle case. The MINLP solver is extremely slow, even in this simplified setting with fixed $S$. The MIQP solver is much faster instead, but still orders of magnitude slower than the NLP solvers. It is hence clear how mixed-integer formulations cannot tackle realistic scenarios in a reasonable time, and the continuous formulation is the only viable option for real-time intersection coordination.

\subsection{Swapping-based heuristic}
\begin{figure*}[!t]
	\centering
	\begin{minipage}[b]{\columnwidth}
		\centering
		\includegraphics[width=\linewidth]{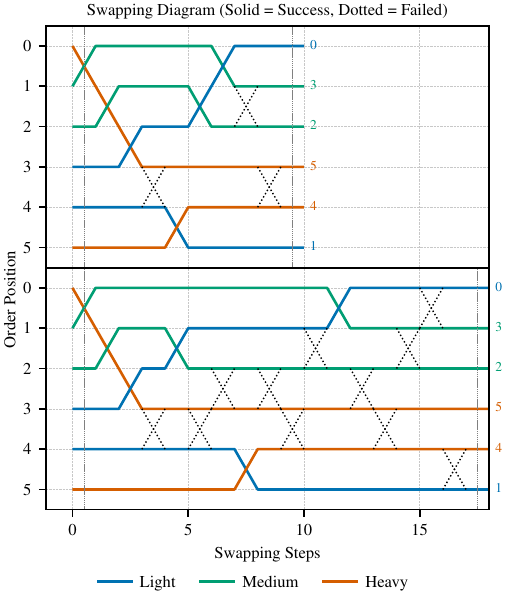}
		\caption{Swapping diagram for 6 vehicles under Heuristic 1 (top) and Heuristic 2 (bottom). The vehicle order is updated based on Lagrange multiplier sensitivity. Solid lines indicate successful swaps; dotted lines are failed attempts.}
		\label{fig:swapping6}
	\end{minipage}\hfill
	\begin{minipage}[b]{\columnwidth}
		\centering
		\includegraphics[width=\linewidth]{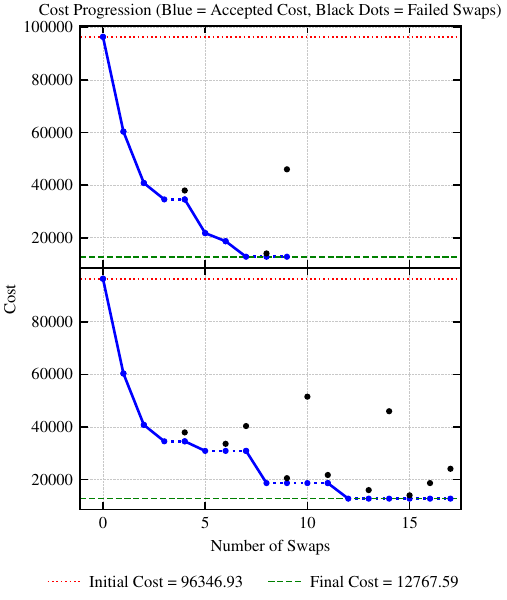}
		\caption{Cost progression during vehicle reordering (6 vehicles) under Heuristic 1 (top) and Heuristic 2 (bottom). The blue line indicates successful swaps with cost reduction; black dots are failed attempts.}
		\label{fig:cost_progression6}
	\end{minipage}
\end{figure*}

\begin{figure*}[t]
	\centering
	\begin{minipage}[b]{\columnwidth}
		\centering
		\includegraphics[width=\linewidth]{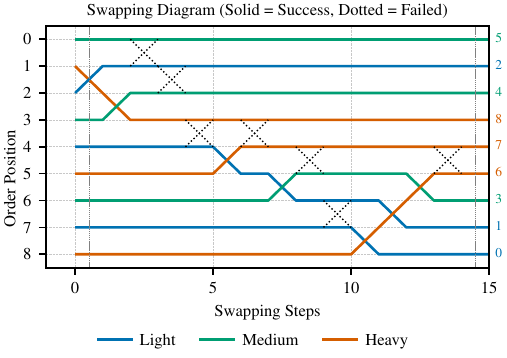}
		\caption{Swapping diagram for 9 vehicles under Heuristic 1. The vehicle order is updated based on Lagrange multiplier sensitivity. Solid lines indicate successful swaps; dotted lines are failed attempts.}
		\label{fig:swapping9}
	\end{minipage}\hfill
	\begin{minipage}[b]{\columnwidth}
		\centering
		\includegraphics[width=\linewidth]{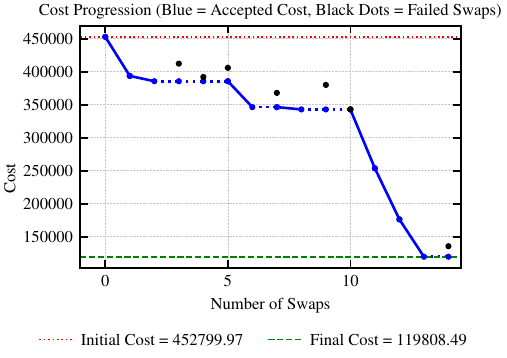}
		\caption{Cost progression during vehicle reordering (9 vehicles). The blue line indicates successful swaps with cost reduction; black dots are failed attempts.}
		\label{fig:cost_progression9}
	\end{minipage}
\end{figure*}

To overcome the intrinsic limitations of FIFO ordering, we apply the Lagrange-multiplier-based swapping heuristic.
Figures~\ref{fig:swapping6} and~\ref{fig:cost_progression6} illustrate the performance of Heuristic 1 (top) and Heuristic 2 (bottom) on a 6-vehicle scenario. The sequence progression is depicted through both successful and failed swap attempts, with the corresponding cost decrease shown across iterations. Similarly, Figures~\ref{fig:swapping9} and~\ref{fig:cost_progression9} illustrate the results for a 9-vehicle case.

% Figs. 4-5: 9-vehicle case

To evaluate the effectiveness of our swapping-based heuristic, we now compare the performance of two NLP solvers, IPOPT and WORHP, on five problem instances. Each instance is characterized by a distinct randomized initial vehicle crossing sequence to which we apply Algorithm~\ref{alg:lambda_swap}.
Table~\ref{tab:solver_comparison} reports the numerical results obtained in a 6-vehicle intersection setting. The initial and final (swapped) orders are shown alongside solver runtimes and total cost reduction percentages. While both IPOPT and WORHP provide the same final reordering and cost, it is clear that the former largely outperforms the latter in terms of computational time. The same trend can also be appreciated in the 9-vehicle case, reported in Table~\ref{tab:solver_comparison_large}.

\begin{table*}[tbh]
	\centering
	\caption{Comparison of IPOPT and WORHP solvers for the 6-vehicle scenario}
	\label{tab:solver_comparison}
	
	\begin{tabularx}{\textwidth}{
			>{\hsize=0.45\hsize}X
			>{\hsize=0.56\hsize}X
			>{\hsize=0.56\hsize}X
			>{\hsize=1.69\hsize}X
			>{\hsize=1.69\hsize}X
			>{\hsize=1.01\hsize}X
			>{\hsize=1.01\hsize}X
			>{\hsize=1.03\hsize}X
		}
		\toprule
		Inst. & IPOPT (s) & WORHP (s) &
		Initial order & Final order &
		Initial cost & Final cost & Reduction (\%) \\
		\midrule
		1 & 16.26 & 35.21 & [2,4,0,1,3,5] & [0,1,2,3,4,5] &
		67431.96 & 16715.55 & 75.21\% \\
		2 & 8.56  & 14.34 & [5,3,1,4,2,0] & [1,3,5,4,2,0] &
		49800.19 & 9060.13  & 81.81\% \\
		3 & 7.37  & 17.28 & [1,4,2,0,5,3] & [1,0,2,4,5,3] &
		48223.00 & 9842.01  & 79.59\% \\
		4 & 14.03 & 23.71 & [5,3,2,0,1,4] & [0,3,2,5,4,1] &
		96346.93 & 12767.59 & 86.75\% \\
		5 & 8.35  & 19.86 & [2,1,3,5,4,0] & [1,2,3,5,4,0] &
		20848.18 & 15266.23 & 26.77\% \\
		\bottomrule
	\end{tabularx}
\end{table*}

\begin{table*}[tbh]
	\centering
	\caption{Comparison of IPOPT and WORHP solvers for the 9-vehicle scenario}
	\label{tab:solver_comparison_large}
	
	\begin{tabularx}{\textwidth}{
			>{\hsize=0.45\hsize}X
			>{\hsize=0.56\hsize}X
			>{\hsize=0.56\hsize}X
			>{\hsize=1.69\hsize}X
			>{\hsize=1.69\hsize}X
			>{\hsize=1.01\hsize}X
			>{\hsize=1.01\hsize}X
			>{\hsize=1.03\hsize}X
		}
		\toprule
		Inst. & IPOPT (s) & WORHP (s) &
		Initial order & Final order &
		Initial cost & Final cost & Reduction (\%) \\
		\midrule
		1 & 56.65 & 96.25  & [8,3,7,2,5,6,4,1,0] & [3,8,7,6,5,4,2,1,0] &
		261533.76 & 72179.17  & 72.40\% \\
		2 & 107.55 & 291.55 & [5,8,2,4,1,7,3,0,6] & [5,2,4,8,7,6,3,1,0] &
		452799.97 & 119808.49 & 73.54\% \\
		3 & 87.11 & 91.61  & [6,0,3,7,8,2,4,1,5] & [0,3,6,7,8,4,5,2,1] &
		248303.87 & 93611.71  & 62.30\% \\
		4 & 93.00 & 358.77 & [4,2,8,0,3,7,1,6,5] & [4,2,0,8,7,6,3,5,1] &
		330573.35 & 129685.82 & 60.77\% \\
		5 & 31.86 & 59.90  & [1,4,7,3,8,6,5,2,0] & [1,4,3,7,8,6,5,2,0] &
		167475.01 & 156774.43 & 6.39\% \\
		\bottomrule
	\end{tabularx}
\end{table*}

\subsection{Comparing swapping heuristics, brute force, and FIFO}
\label{sec:comparative_evaluation}
We evaluate next our Lagrange-multiplier-based reordering strategy and compare it to brute-force search and FIFO scheduling, across 100 different scenarios with 6 vehicles.

\begin{figure}[t]
    \centering
    \includegraphics[width=\columnwidth]{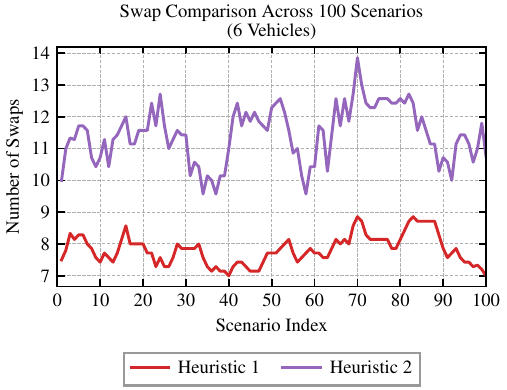}
    \caption{Number of swaps across 100 scenarios for Heuristic 1 and 2 with 6 vehicles. 
    }
    \label{fig:swap_comparison}
\end{figure}

Brute-force search provides a globally optimal benchmark by evaluating all admissible vehicle orderings. However, the number of possible schedules, $M!$, grows combinatorially with the number of vehicles,
making the exhaustive enumeration intractable even for moderate problem sizes. While both heuristics require only a small fraction of the NLP solves needed by brute-force search, they are in fact equivalent in solution quality: across all scenarios, Heuristic~1 and Heuristic~2 return the same final crossing order, and hence the same cost. They differ only in computational effort, that is, in the number of NLPs that must be solved to reach that order. As empirically confirmed in Figure~\ref{fig:swap_comparison}, Heuristic 1 requires fewer swap evaluations across all scenarios, owing to its restricted search space controlled by the parameter $\alpha$ in Algorithm~\ref{alg:lambda_swap}, and is therefore the more efficient of the two. Because the two heuristics returned the same solution in all simulations we ran, we only report the results of Heuristic 1 for the remaining experiments. 

Then, let \(J_{\mathrm{BF}}^{(s)}\), \(J_{\mathrm{FIFO}}^{(s)}\), and \(J_{\mathrm{H}}^{(s)}\) denote the total cost of brute force, FIFO, and Heuristic~1, respectively, in scenario \(s\). The optimality gap reduction (OGR) is
\begin{equation}
\mathrm{OGR}^{(s)} =
\frac{J_{\mathrm{FIFO}}^{(s)} - J_{\mathrm{H}}^{(s)}}{J_{\mathrm{FIFO}}^{(s)} - J_{\mathrm{BF}}^{(s)}} \times 100,
\end{equation}
and the optimality gap (OG) is
\begin{equation}
\mathrm{OG}^{(s)} =
\frac{J_{\mathrm{H}}^{(s)} - J_{\mathrm{BF}}^{(s)}}{J_{\mathrm{BF}}^{(s)}} \times 100.
\end{equation}

\begin{figure}[t]
    \centering
    \includegraphics[width=\columnwidth]{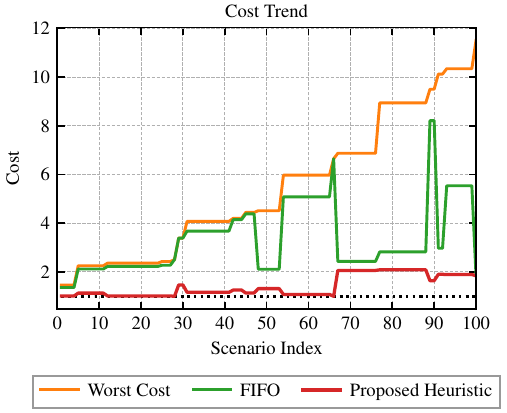}
    \caption{Cost trend across 100 scenarios for brute force, FIFO, and the proposed heuristic. All costs are normalized by the brute-force (optimal) cost, shown as the black dotted line, and scenarios are sorted by the worst-case cost.}
    \label{fig:cost_comparison}
\end{figure}

In Figure~\ref{fig:cost_comparison} we illustrate the cost trend across all scenarios. As expected, brute force provides the optimal lower bound, while FIFO yields significantly higher costs. Compared to FIFO, our heuristic scheme consistently reduces the cost and follows the optimal trend in most cases.

\begin{figure}[t]
    \centering
    \includegraphics[width=\columnwidth]{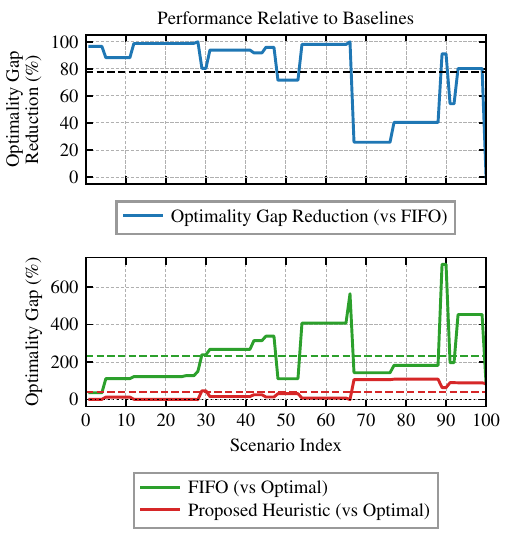}
    \caption{Performance metrics across 100 scenarios. Top: optimality gap reduction relative to FIFO. Bottom: optimality gap of FIFO and our proposed heuristic relative to the brute-force optimal solution. 
    }
    \label{fig:performance_metrics}
\end{figure}

Figure~\ref{fig:performance_metrics} reports both OGR and OG. Our heuristic achieves an average OGR of approximately \(77.4\%\) relative to FIFO, with several scenarios approaching near-complete gap closure. At the same time, the average OG relative to the brute-force optimal solution is around \(41.8\%\), indicating a practical trade-off between solution quality and computational efficiency.

\subsection{12-Vehicle scenario: Scalability analysis}
\begin{figure}[t]
\centering
\includegraphics[width=\columnwidth]{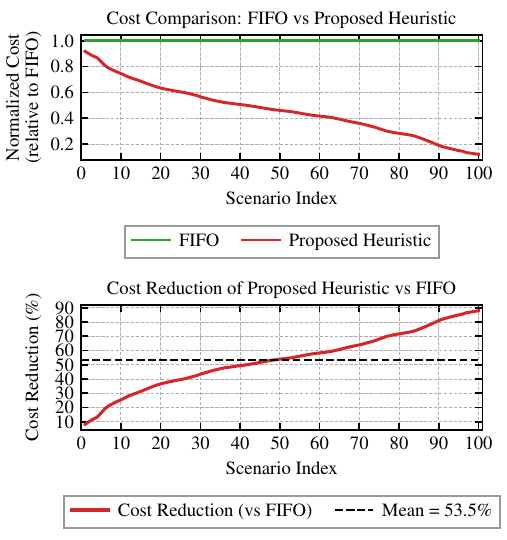}
\caption{Cost comparison for 100 scenarios with 12 vehicles, sorted by cost reduction. Top: normalized cost relative to FIFO; 
Bottom: cost reduction of the proposed heuristic relative to FIFO.}
\label{fig:cost_comparison12}
\end{figure}

\begin{figure}[t]
\centering
\includegraphics[width=\columnwidth]{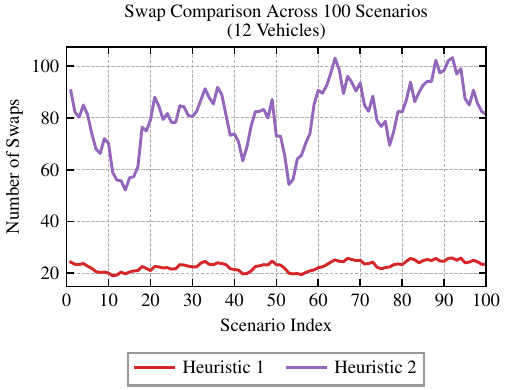}
\caption{Number of swap evaluations per scenario for Heuristic 1 and Heuristic 2 with 12 vehicles.}
\label{fig:swap_comparison12}
\end{figure}
To assess the scalability of the proposed heuristic beyond the six-vehicle case, we conduct a large-scale evaluation on 100 randomly generated scenarios with \(M = 12\) vehicles. With this problem size, brute-force enumeration becomes computationally infeasible, since the number of admissible schedules is \(12! = 479{,}001{,}600\). Consequently, FIFO scheduling serves as the sole reference baseline, and our objective is to quantify the improvement achieved by the proposed Lagrange-multiplier-based reordering strategy relative to this baseline.
Figure~\ref{fig:cost_comparison12} reports the normalized cost of the proposed heuristic relative to FIFO across all 100 scenarios, sorted by increasing cost reduction, together with the corresponding cost reduction. The heuristic consistently achieves costs well below the FIFO baseline: the normalized cost ranges from values close to unity in the least favorable scenarios down to approximately 0.1 in the most favorable ones, confirming that meaningful improvements are obtained across the entire range of scenarios. The average cost reduction relative to FIFO is approximately \(53.5\%\), with individual scenarios exhibiting reductions ranging from roughly \(9\%\) up to nearly \(88\%\). The distribution of improvements is broadly spread across the scenario set, indicating that the heuristic is not limited to favorable instances but consistently exploits scheduling structure across heterogeneous traffic conditions.

Figure~\ref{fig:swap_comparison12} compares the two heuristic variants in terms of swap count. Heuristic 1 requires, on average, approximately 20 to 25 swap evaluations per scenario, whereas Heuristic 2 performs substantially more, ranging between 55 and 105 swaps. This difference results from the mechanism of Heuristic 1 that prevents revisiting previously evaluated pairs and thereby produces a non-redundant search. These trends are consistent with those observed in the six-vehicle case and confirm that the computational advantage of Heuristic 1 is preserved as the problem dimension increases.

These results demonstrate that the proposed framework scales favorably to larger vehicle populations. While the absence of a global optimal reference precludes direct optimality gap quantification at this scale, the consistent and substantial reductions over FIFO, paired with the fast search of Heuristic 1, confirm that our schedule refinement remains an effective and efficient coordination strategy as the combinatorial complexity of the scheduling problem grows.

\subsection{Comparison against rule-based and reservation-based heuristics}
\label{sec:baseline_comparison}
\begin{figure}[t]
    \centering
    \includegraphics[width=\columnwidth]{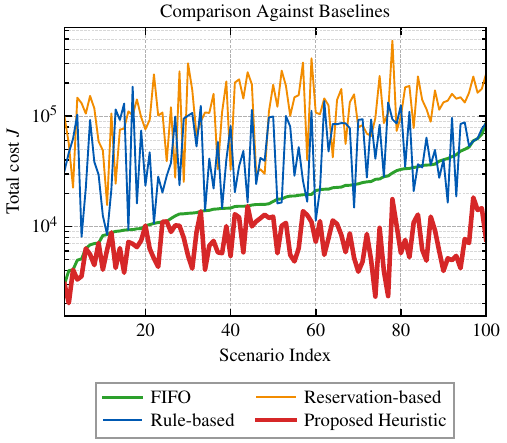}
    \caption{Total cost across 100 randomized 6-vehicle scenarios, sorted by FIFO cost. The rule-based baseline and the reservation-based scheme exceed even the FIFO cost, while the proposed heuristic attains the lowest cost in all scenarios.}
    \label{fig:baseline_comparison}
\end{figure}
Finally, we compare the proposed heuristic against representatives of the two classes of computationally light coordination schemes discussed in the introduction, over the same 100 randomized 6-vehicle scenarios considered in Section~\ref{sec:comparative_evaluation}. Rule-based policies fix the crossing order through predefined priorities, independently of the vehicles' objectives~\cite{qian2014priority}. As a representative of this class, we evaluate a static right-of-way convention in which main-road approaches take precedence over secondary ones, with vehicles ordered by arrival within each class. The reservation-based baseline implements the first-come-first-served scheme of~\cite{dresner2008multiagent}: a central manager processes vehicles in arrival order and grants each one the earliest conflict-free occupancy slot no earlier than its free-flow crossing interval, and each vehicle then plans its trajectory individually so as to comply with the granted slot. In both cases, the resulting schedules are evaluated with the objective of~\eqref{eq:coordination_ocp}, so that all methods are directly comparable. Figure~\ref{fig:baseline_comparison} reports the total cost across all scenarios. Both baselines are consistently outperformed: on average, the right-of-way rule is approximately $2.7\times$ and the reservation-based
scheme approximately $6\times$ more expensive than FIFO, as cost-blind ordering and rigid slot allocation, respectively, penalize precisely the vehicles with the largest weights. We remark that these schemes target throughput and provable collision avoidance rather than the objective of~\eqref{eq:coordination_ocp}, and are considerably cheaper to evaluate; the comparison thus quantifies the performance they forgo when vehicles have heterogeneous objectives. The proposed heuristic attains the lowest cost in all 100 scenarios, with an average cost reduction of $51.2\%$ with respect to FIFO.

Although the proposed heuristic is orders of magnitude faster than the mixed-integer formulations, its runtime is not yet compatible with real-time operation, since the schedule is recomputed from scratch for a fixed set of vehicles and every candidate swap requires a full NLP solve. Both aspects can be improved: the schedule can be updated incrementally as vehicles arrive, within a receding-horizon scheme, and the NLP re-solves can be replaced by sensitivity-based approximations that exploit the bi-level structure of the problem.

\section{Conclusion}
We have presented an optimization-based framework for coordinating autonomous vehicles at unsignalized intersections.

Having formulated the overall control problem as a nonlinear program, we have proposed a Lagrange-multiplier-based heuristic to address the combinatorial nature of the vehicles crossing order scheduling. By exploiting sensitivity information available for free from the KKT conditions, our technique refines vehicle crossing orders through targeted swaps, avoiding exhaustive enumeration or combinatorial optimization. 

Evaluated against brute-force search, FIFO scheduling, and mixed-integer formulations on extensive numerical simulations, our scheme has demonstrated several advantages. Specifically, it usually requires significantly lower computational time compared to a mixed-integer setup and brute-force search, while achieving a comparable performance in terms of cost minimization, and improving crossing order over the basic FIFO approach. The proposed architecture then enables efficient approximation of optimal schedules while preserving computational tractability, making it suitable for real-time intersection coordination.

Future work will extend the framework in two directions that build on each other. First, the heuristic can be embedded in a receding-horizon scheme in which vehicles enter the coordination area over time, so that the crossing order is updated incrementally at each arrival rather than recomputed from scratch. Second, the computational cost of the swap evaluations can be reduced by exploiting the bi-level structure of the problem and the parametric sensitivities of the vehicle trajectory problems, with the aim of executing the reordering within the sampling time. Further directions include robustness to model uncertainty, multi-intersection coordination, and integration with learning-based components for adaptive decision-making.

% \clearpage

\bibliographystyle{IEEEtran}
\bibliography{references}

\end{document}